\documentclass[aoas,nameyear,dvips]{arximspdf}

\doi{10.1214/09-AOAS312G}
\referstodoi{10.1214/09-AOAS312}
\volume{3}
\issue{4}
\pubyear{2009}
\firstpage{1299}
\lastpage{1302}



\begin{document}
\begin{frontmatter}

\title{Discussion of: Brownian distance covariance}
\runtitle{Discussion}
\pdftitle{Discussion on Brownian distance covariance by G. J. Szekely and M. L. Rizzo}

\begin{aug}
\author{\fnms{Christopher R.} \snm{Genovese}\corref{}\ead[label=e1]{genovese.cr@gmail.com }}
\runauthor{C. R. Genovese}
\affiliation{Carnegie Mellon University}
\end{aug}

% HISTORY:

% ABSTRACT

% KEYWORDS

\end{frontmatter}

Congratulations to Professors Sz\'{e}kely and Rizzo for such an
exciting and enjoyable contribution.
It is not often that one of our most basic techniques is given so
fundamental, and so successful, a rethinking.
Although using distance covariance requires giving up some useful
properties associated
with linearity---directionality/sign, exact expressions for the
variance and covariance of sums,
direct connection to the multivariate normal distribution---it offers
useful properties in exchange.
Distance covariance gives a true indicator of independence even for
non-normal distributions,
applies directly in multivariate settings (even when ``$p \gg n$''),
is the basis for general and powerful tests,
can be adapted to use ranks,
provides conditions for central limit theorems,
and is straightforward to compute.
That seems to be a favorable trade.
In this discussion I will focus on the meaning of Brownian covariance,
but first I want to raise a few questions to the authors (and the field).

The paper adapts the statistic in examples to derive resampling techniques
and tests for nonlinearity and extends the covariance definition in
several ways.
Perhaps the authors can comment on how general these derived techniques are.
For instance, what additional conditions, if any, are required for the test
of nonlinearity in Example 6 (based on $\mathsf{dCov}(X, (I - X (X^T
X)^{-1} X^T) Y)$)
to be consistent?
Also, the computations would appear to be $O(n^2)$, which can be
burdensome for very large $n$.
Are there speed-ups or approximations that yield comparable results
more quickly?
And are rates of convergence available for the empirical statistics,
perhaps under stronger moment conditions?

But these are details. Even though the Pearson correlation is
entrenched in the practice
of several fields, including our own,
what reason do we have not to aggressively introduce distance
covariance and correlation
into our practice \emph{and} our teaching, even at the introductory level?
It is rare in practice that we want a measure of linear association
\emph{per se},
more typically we use Pearson correlation as a proxy.
Distance covariance provides most of what we do want in these cases
with attendant theory and convenience
that is hard to beat.
And teaching about the difference between ``uncorrelated'' and
``independent'' is a thorn in the side
of anyone who has had to do so. Distance covariance would require no
more sophisticated ideas
than what we already use in teaching correlation, without that complication.
The statistic is expressed in terms of distances which are easy to understand,
and it would free us from undue emphasis on Normal examples.
It is interesting to ponder what it would take to change practice at
this level.

However, what principally distinguishes this paper from Sz\'{e}kely,
Rizzo and Bakirov (\citeyear{2007})
is the introduction of Brownian covariance.
Because of the ``suprising coincidence'' that Brownian covariance
equals distance covariance,
though under slightly more restrictive conditions,
Brownian covariance may appear to be merely an interesting, if
abstract, representation.
But Brownian covariance can help us understand how and why distance
covariance works
and how it can be generalized to obtain measures with other desirable
properties.
As the authors write, Brownian covariance ``measures the degree of
\emph{all kinds
of possible relationships} between two real-valued random variables.''
Because it may not be obvious why this statement is true,
my goal here is to explain it in a different way
and to offer insight into what the Brownian covariance means and how it
can be usefully
generalized.
I will do this by studying the $(U,V)$-covariance (Definition~5 in the
paper) for a special class of stochastic
processes.
To keep the focus on the ideas rather than details,
I will consider only a simple case here, and I will play somewhat loose
with regularity conditions (e.g., limit interchanges),
but all of this can be made rigorous and general without excessive
effort or conditions.\looseness=1

Let $X$ and $Y$ be scalar random variables with finite second moments.
Denote their joint density by $g_{X,Y}$ and marginal densities by $g_X$
and $g_Y$.
For simplicity, assume that these densities are square integrable and
have support
on $[0,1]^2$ and $[0,1]$ respectively, although these restrictions can
easily be weakened.
Let $(\phi_i)$ and $(\psi_j)$ be two sequences of deterministic functions.
They may be finite or infinite collections and need not be orthogonal.
Define
\begin{eqnarray}
A_{ij} &=& \int\!\!\int(\phi_i \otimes\psi_j)\, (g_{X,Y} -
g_X\otimes g_Y)
\\
&=& \int\!\!\int(\phi_i \otimes\psi_j)\, g_{X,Y} - \int\phi_i\,
g_X \int\psi_j\, g_Y
\\
&=& \mathsf{Cov}(\phi_i(X), \psi_j(Y))
\\
&=& \mathbb{E} X_{\phi_i} Y_{\psi_j}.
\end{eqnarray}
Now consider stochastic processes $U$ and $V$ that can be written as
series expansions
with Normal coefficients.
Specifically, given suitable positive values $\sigma_i$ and $\tau_j$, define
\begin{eqnarray}
U(s) &=& \sum_i \sigma_i Z_i \phi_i(s),
\\
V(t) &=& \sum_j \tau_j Z'_j \psi_j(t),
\end{eqnarray}
where the $Z_i$'s and $Z'_j$'s are independent standard Normal random variables.

Using the notation of the paper and
interchanging expectation and sums in the definitions of $X_U = U(X) -
\mathbb{E}(U(X) \mid U)$ and related random
variables, we have
\begin{eqnarray*}
X_U &=& \sum_i \sigma_i Z_i \bigl(\phi_i(X) -
\mathbb{E}\phi_i(X)\bigr),
\\
X'_U &=& \sum_i \sigma_i Z_i \bigl(\phi_i(X') -
\mathbb{E}\phi_i(X')\bigr),
\\
Y_V &=& \sum_j \tau_j Z'_j \bigl(\psi_j(Y) - \mathbb{E}\psi_j(Y)\bigr),
\\
Y'_V &=& \sum_j \tau_j Z'_j \bigl(\psi_j(Y') - \mathbb{E}\psi_j(Y')\bigr).
\end{eqnarray*}
%
%Define $S_{ij} = \sigma_i^2 \delta_{ij}$ and $T_{ij} = \tau_j^2
%for Kronecker delta $\delta_{ij}$.
%and for convenience treat $A$, $S$, and $T$ as (possibly infinite
%order) matrices.
It follows from Definition 5 in the paper that
\begin{eqnarray}\label{eq::covexpr}
\mathsf{Cov}_{U,V}^2(X,Y)
&=& \mathbb{E}(X_U X_U' Y_V Y'_V) \nonumber
\\
&=& \sum_{i,j,k,\ell} \mathbb{E}(Z_i Z_k)\, \mathbb{E}(Z'_j Z'_\ell) \,\mathbb{E}( X_{\phi
_i} Y_{\psi_j} ) \,\mathbb{E}( X_{\phi_k} Y_{\psi_\ell} )
\\
&=& \sum_{i,j} \sigma_i^2 \tau_j^2 A_{ij}^2.\nonumber
\end{eqnarray}
%
%where $\norm{M}_F = \sqrt{\sum_{i,j} M_{i,j}^2}$ is the Frobenius
%matrix norm.
Equation (\ref{eq::covexpr}) shows that $\mathsf{Cov}_{U,V}(X,Y) = 0$
if and
only if every $A_{ij} = 0$.
For this covariance to determine independence, we must have that all
$A_{ij} = 0$ if and only if
$X$ and $Y$ are independent.
A sufficient condition for this is that the functions $\phi_i \otimes
\psi_j$ form a (Schauder) basis for
a class of functions containing $g_{X,Y} - g_X \otimes g_Y$ (e.g.,
$\mathcal{L}^2$).
Note that in this case
\begin{eqnarray}
(f_{X,Y} - f_X \otimes f_Y)(s,t)
&=& \int\!\!\int e^{i (s x + t y)} (g_{X,Y} - g_X \otimes g_Y)(x,y)\,
{\rm d} x\,{\rm d} y \\
&=& \int\!\!\int e^{i (s x + t y)} \sum_{i,j} A_{ij} \phi_i(x) \psi
_j(y) \,{\rm d} x\,{\rm d} y \\
&=& \sum_{ij} A_{ij} \tilde\phi_i(s) \tilde\psi_j(t),
\end{eqnarray}
where the $f$'s are the characteristic functions of the $g$'s as in the paper
and the $\tilde\phi$'s and $\tilde\psi$'s are the corresponding
Fourier transforms.
This shows that the covariance is related to a ``norm'' of $f_{X,Y} -
f_X \otimes f_Y$
and thus highlights the connection to the distance covariance as
defined in the paper.

Now it is well known (the L\'{e}vy--Ciesielski construction) that
Brownian motion can be written as
\begin{equation}\label{eq::LCexpansion}
W_t = \sum_{i\ge0} Z_i S_i(t),
\end{equation}
where $S_i$ is the $i$th Schauder function obtained by $S_i(t) = \int
_0^t H_i$
for the corresponding function $H_i$ in the Haar system.\footnote{This
construction is usually shown for $t\in[0,1]$ but can be extended recursively;
for instance, if $t\in(1,2]$, define $W_{t} = W_1 + \sum_{i\ge0}
Z_i'' S_i(t-1)$
with independent $Z''_i$s. Using the full line would require a slightly
more general form
of equation (\ref{eq::covexpr}), which is straightforward to derive.}
The expansion (\ref{eq::LCexpansion}) corresponds to $U = W$ and $V = W'$
with $\sigma_k = \tau_k = 1$ and $\phi_k = \psi_k$ equal to
corresponding Schauder functions for all $k$.
Hence,
\begin{equation}
\mathsf{Cov}_W(X,Y) = \sqrt{\sum_{i,j} A_{ij}^2},
\end{equation}
the Frobenius norm of the infinite-order matrix $A$.
Because the Schauder functions form a (nonorthogonal) basis for the set
of continuous functions on an interval (in sup-norm)
and for the $\mathcal{L}^p$ spaces for $1 < p < \infty$,
we can see that a zero Brownian covariance is equivalent to
independence of $X$ and $Y$.
Because the Schauder functions have support in nested (and shrinking)
dyadic intervals,
$A_{ij}^2$ measures the dependence in $g_{XY}$ over a small dyadic rectangle.
The Brownian covariance thus combines measures of dependence across all
scales in
a multi-resolution hierarchy,
and this is the sense in which it captures all kinds of dependence.
This derivation also clarifies how changing the stochastic processes
$U$ and $V$ can
give covariance measures that emphasize different features of $X$ and
$Y$'s joint distribution.


\begin{thebibliography}{99}

\bibitem[\protect\citeauthoryear{}{2007}]{2007}
\textsc{Sz\'{e}kely, G. J.}, \textsc{Rizzo, M. L.} and \textsc
{Bakirov, N. K.}
(2007). {Measuring and testing dependence by correlation of distances}.
\textit{Ann. Statist.} \textbf{35} 2769--2794.
\MR{2382665}

\end{thebibliography}
\end{document}